\documentclass{article}
\usepackage{spconf,amsmath,graphicx}

\usepackage{xcolor}
\usepackage{placeins}

\newcommand*{\com}{\textcolor{blue}}

\usepackage{enumitem}
\setlist{nosep, leftmargin=14pt}

\usepackage{mwe} 

\usepackage{adjustbox}
\usepackage{numprint}
\usepackage{multirow}
\usepackage{booktabs}

\title{Disentanglement enables cross-domain Hippocampus Segmentation}
\name{John Kalkhof, Camila Gonz\'{a}lez, Anirban Mukhopadhyay}
\address{Darmstadt University of Technology, Karolinenplatz 5, 64289 Darmstadt, Germany}
\begin{document}
%
\npdecimalsign{.}
\nprounddigits{3}

\maketitle

© 2022 IEEE.  Personal use of this material is permitted.  Permission from IEEE must be obtained for all other uses, in any current or future media, including reprinting/republishing this material for advertising or promotional purposes, creating new collective works, for resale or redistribution to servers or lists, or reuse of any copyrighted component of this work in other works.

\begin{abstract}

Limited amount of labelled training data are a common problem in medical imaging. This makes it difficult to train a well-generalised model and therefore often leads to failure in unknown domains. Hippocampus segmentation from magnetic resonance imaging (MRI) scans is critical for the diagnosis and treatment of neuropsychatric disorders. Domain differences in contrast or shape can significantly affect segmentation. We address this issue by disentangling a T1-weighted MRI image into its content and domain. This separation enables us to perform a domain transfer and thus convert data from new sources into the training domain. This step thus simplifies the segmentation problem, resulting in higher quality segmentations. We achieve the disentanglement with the proposed novel methodology 'Content Domain Disentanglement GAN', and we propose to retrain the UNet on the transformed outputs to deal with GAN-specific artefacts. With these changes, we are able to improve performance on unseen domains by 6-13\% and outperform state-of-the-art domain transfer methods. 

\end{abstract}
\begin{keywords}
feature disentanglement, domain generalisation, distribution shift 
\end{keywords}
\section{Introduction}
\label{sec:intro}
In medical imaging, it is common that only a limited amount of diverse labelled data is available for training. This is problematic as the availability of data from different domains is crucial to train a well-generalised model \cite{memmel2021adversarial}. Without proper generalisation, it is likely that segmentation fails when dealing with out-of-distribution data \cite{sanner2021reliable}. Segmentation of the hippocampus based on MRI scans is particularly important for the diagnosis and treatment of neuropsychiatric disorders \cite{sanner2021reliable}. However, only a few annotated datasets are available for this type of data. This confronts us with the aforementioned problem of domain differences. 

Feature disentanglement enables us to transfer knowledge and consequently create more robust models. By disentangling the domain-specific features from the content of the image we ensure that the segmentation quality of the model is not affected by domain differences. This is an advantage compared to other state-of-the-art methods for hippocampus segmentation \cite{LIU2020116459, Cao2018, Ataloglou2019} as they cannot be optimised to process data from unseen sources without domain-specific labelled data.

This setup requires a model that achieves \textbf{strong disentanglement} to completely separate content and domain-specific features. It needs to be able to \textbf{transform data from different domains into the same space}, to remove domain-specific features and create a well-generalised representation. Finally, it needs \textbf{high-quality segmentation} results on this disentangled image representation.

\begin{table}[htb]
    \centering
    \begin{adjustbox}{max width=8.5cm}
    \begin{tabular}{cccc}
        \toprule
        & \textbf{Strong} & \textbf{Domain Transfer} & \textbf{High Quality} \\
        & \textbf{Disentanglement} & \textbf{Capability} & \textbf{Segmentation} \\
        \midrule
        \textbf{SA-GAN \cite{semantic_aware2018}} & $\boldsymbol{--}$ & $\boldsymbol{++}$ & $\boldsymbol{++}$ \\
        \textbf{SD-Net \cite{sdnet2019}} & $\boldsymbol{+-}$ & $\boldsymbol{+-}$ & $\boldsymbol{++}$ \\
        \textbf{Dr-GAN \cite{Tran_2017_CVPR}} & $\boldsymbol{++}$ & $\boldsymbol{++}$ & $\boldsymbol{--}$ \\
        \textbf{CDD-GAN (ours)} & $\boldsymbol{++}$ & $\boldsymbol{++}$ & $\boldsymbol{++}$ \\
        \bottomrule
    \end{tabular}
    \end{adjustbox}
    \caption{Comparison of state-of-the-art approaches based on our requirements.}
    \label{tab:requirements}
\end{table}

We investigate several methodologies (see Table \ref{tab:requirements}) as a basis for our work. Although none of these methodologies can fully meet our requirements, Dr-GAN is the most promising as it achieves strong disentanglement and is able to perform transformations between disentangled variables. Since Dr-GAN is developed with classification in mind, the disentangled output can differ strongly from the input, which is problematic when dealing with segmentation tasks.

In this work, we present the novel method \textbf{'Content Domain Disentanglement GAN' (CDD-GAN)}. Our method consists of two main parts: first, the disentanglement network, which generates a better-generalised representation for input images from different sources. Second, the segmentation network, which is trained on the output of the generator. The disentanglement network is based on the structure of Dr-GAN, but improves its output by adding a cycle consistency loss. This loss directly optimises the models ability to recreate the original input image. By forcing the model to create a latent representation that stores all the information to recreate the image, the shape of the outputs becomes more similar to the input and is much better suited for high-quality segmentation. 

For the second part, we propose to retrain a UNet \cite{unet2015} on the outputs of Dr-GANs generator $G$. Due to the bottleneck in the latent representation, the outputs of GANs generally do not match the input when they are recreated, and small differences to the input image are lost. This is problematic because the pretrained UNet is optimised on the raw input images. The solution we propose is to transform our original input images by the generator $G$ into the same domain and then train our UNet on these outputs. This way we do not need any additional labelled data, but our newly trained UNet is now optimised to deal with the outputs of the generator.

For the evaluation, we create two distorted datasets by synthetically adding realistic distortions to test the robustness of the proposed CDD-GAN. We randomly change either the contrast or elastically deform the original datasets images. The use of synthetically distorted datasets is crucial for gaining insight into the domain differences and allows us to quantitatively evaluate disentanglement. Furthermore, it helps to avoid silent failures as it allows a direct comparison between transformation results.

Lastly, we perform a thorough comparison with Dr-GAN \cite{Tran_2017_CVPR} as well as two other state-of-the-art methodologies: SA-GAN \cite{semantic_aware2018} and SD-Net \cite{sdnet2019}. SA-GAN is a domain-transfer approach that learns a one-to-one transformation between domains. SD-Net is based on VAEs \cite{kingma2014autoencoding, Higgins2017betaVAELB} and learns to segment by disentangling into content and domain and then segments only on the latent variables of the content.

The contributions made in this work are the following:
\begin{itemize}
 \renewcommand{\labelitemi}{$\bullet$}
 \item We propose CDD-GAN, which adapts the Dr-GAN method for disentanglement in content and domain and adds a cycle loss component to produce more accurate transformed images for segmentation. Further, we combine it with a UNet that is optimised on the generalised outputs of the generator.
 \item We introduce an evaluation scheme suitable for testing the performance of disentangling-based segmentation methods, which consists of two distorted datasets.
\end{itemize}
We make our code publicly available at \textit{anonymized repository}, and hope that our contribution pushes forward the structured evaluation of disentanglement methods for medical imaging segmentation.

\section{Materials and Methods}
\label{sec:methods}
In this section we outline how we adapt Dr-GAN to the segmentation task. We then introduce our new methodology CDD-GAN. Finally, we discuss the advantages of retraining the UNet on the outputs of the generator.

\subsection{Dr-GAN} \label{sec:meth_drgan}
The \emph{disentangled representation learning GAN} \cite{Tran_2017_CVPR} was originally developed to disentangle the angle of the face from the identity of the subject in order to achieve better face recognition.
In our setup, disentanglement between the content and domain-specific features of the input image is performed. Therefore, Dr-GAN is trained as follows. 

First, an image is fed into the encoder $G_{enc}$, resulting in a latent representation $f(x)$. In addition to these latent variables, the random domain $d_r$ and a noise variable $z$ are added. The noise variable $z$ is used to synthesise an image, where $f(x)$ is the identity and $d$ is the domain. All these variables then go into the decoder $G_{dec}$ to generate an image $\bar{x}$. This image $\bar{x}$ then goes into the discriminator $D$, which not only decides whether the image is real or not, but also identifies the original domain as well as the identity. The model is then trained in the GAN-typical \cite{goodfellow2014generative} two-player game.

\begin{figure}[htb]

\begin{minipage}[b]{1.0\linewidth}
  \centering
  \centerline{\includegraphics[width=8.5cm]{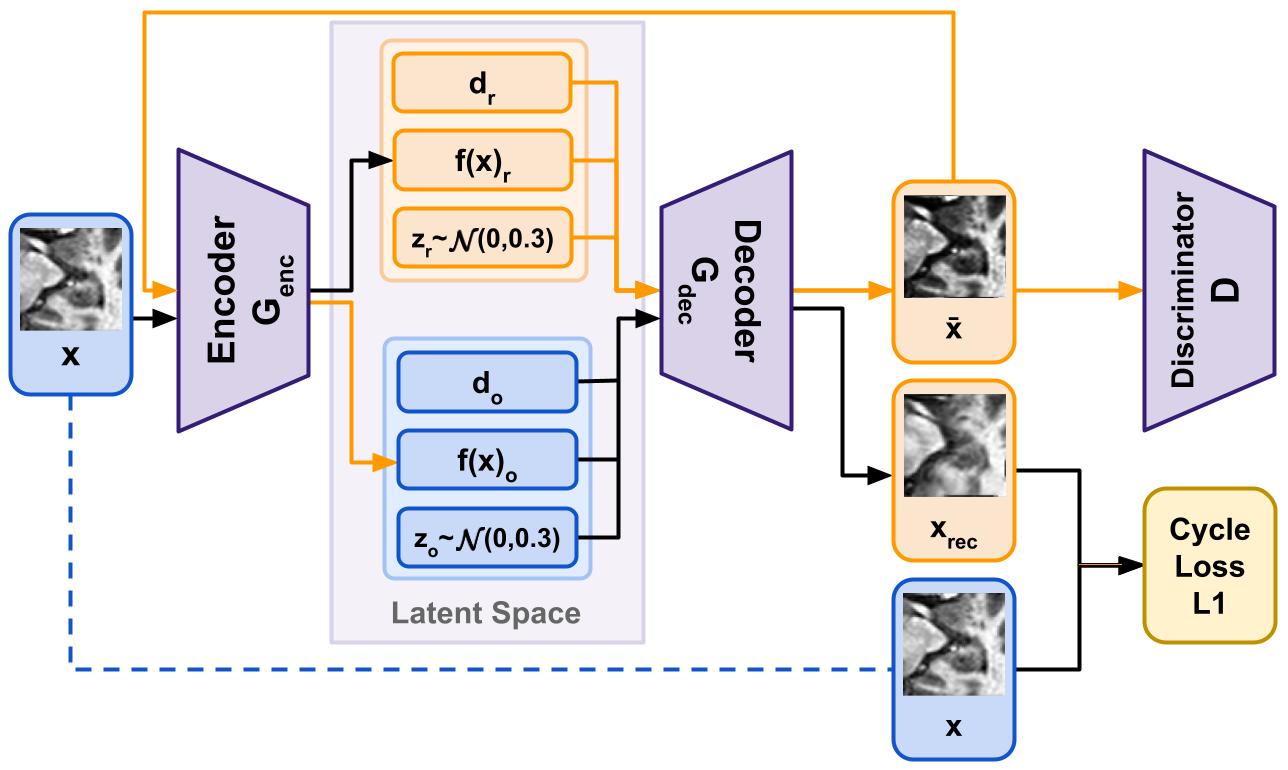}}
\end{minipage}
\caption{CDD-GAN setup using the cycle consistency loss and an identity/domain discriminator for the training.}
\label{fig:CDDGan}
\end{figure}

\subsection{Cycle Consistency Loss} \label{sec:meth_cycle}
When disentangling data into multiple factors, it is not only important that the features are independent, but also that as little information as possible is lost during this process \cite{eastwood2018a}. Dr-GAN solves this problem by simultaneously optimising the discriminator $D$ and the generator $G$ to preserve the identity of the input image $x$ in the output $\bar{x}$. However, this only forces the generator $G$ to provide enough information for the discriminator $D$ to find out the identity. Therefore, it is likely that some of the information necessary to return to the original domain $d$ and restore the image after transforming it into another domain is lost. 

We approach this problem, inspired by the SA-GANs \cite{semantic_aware2018} architecture, by introducing an L1 cycle consistency loss into the Dr-GAN structure. The L1 loss directly optimises the ability to reconstruct the input image from the output $\bar{x}$. The idea behind this approach is that the generator $G$ must now not only convince the discriminator $D$, but also must be able to recover the original image after a random domain transfer. This forces Dr-GAN to store more details about the image that are important for reconstruction and therefore relevant for segmentation. 

The cycle loss quantifies the difference between the reconstructed image $x_{rec}$ and the input image $x$, as illustrated in Figure \ref{fig:CDDGan}. It is implemented as the L1-loss of $x_{rec}$ and $x$. When training the model, the input data is transformed into a random domain $d_r$. The output of this transformation $\bar{x}$ is then passed through the generator again, with the original input domain $d_o$ as the target.

\subsection{UNet Retraining}
\label{sec:retraining}
Due to the bottleneck structure of the generators-encoder-decoder architecture, information is always lost when the output image is generated. This means that the output $\bar{x}$ is almost never equal to the input $x$, even if the same domain is kept. As a result, a pretrained UNet often performs worse on the generator output $\bar{x}$ than on the non-transformed data $x$. 

We propose to retrain the UNet on the output of the generator $G$. This means that we convert all data belonging to the original domain to the target domain $d=O$ using the generator. Thus, we transform the images by the generator but keep the target domain the same as the input. The output image $\bar{x}$ thus corresponds to $x$ plus the GAN-specific artefacts. This approach does not require any additional labelled data, but has the potential to further improve the performance by accounting for the GAN-specific artefacts. 

\section{Experimental Setup}
\label{sec:exp_setup}
In the following chapter we introduce the data as well as the creation of the distorted datasets. We briefly introduce the hardware used and finally discuss our evaluation metrics.

\subsection{Data: Hippocampus-(O/C/E)}
\label{sec:data}
For the evaluation, we use the hippocampus dataset made available by the \textit{Medical Segmentation Decathlon} \cite{medicaldecathlon}. This study contains T1-weighted 3D images from 260 patients. We split the data into sets of 146 patients for training, 62 for validation and 52 for testing. The images range in size from 31x40 to 43x59 pixels and are scaled to slices of 64x64. We refer to the original images as the \textbf{H\textsubscript{O}} (original) dataset. 

We create two distorted datasets based on this dataset. The first, the \textit{contrast} dataset, is scaled by a power function with a random contrast factor between 1 and 2 and is henceforth called \textbf{H\textsubscript{C}}. The second distorted dataset uses TorchIO's \cite{perez-garcia_torchio_2020} elastic deformation function with a grid size of 5x5 and a maximum displacement of 12 to deform the image and is from now on called \textbf{H\textsubscript{E}}. Sample images for all three datasets can be found in the first column of Figure \ref{fig:prediction_comparison}. The use of synthetically distorted datasets is crucial as it gives us insight into the domain-specific differences and therefore allows us to quantify the disentanglement. Further details on how we generate this database can be found in our code base.

\subsection{Hardware}
\label{sec:hardware}
All training is performed on a machine with an AMD Ryzen 5600X processor and an Nvidia RTX 2060 Super GPU (8GB VRAM).

\subsection{Metrics}
\label{sec:metrics}
The evaluation focuses on two major quality factors. The first is the \emph{quality of the actual segmentation}. We use the Dice metric for measuring segmentation performance. For calculating subject-wise Dice scores, the masks are predicted layer by layer and then combined to form a 3D volume. 
The second factor is the \emph{quality of the disentanglement}, which we measure with the \emph{classification score} of the discriminator. 

For each transformed set of hippocampus datasets, we train a ResNet-18 \cite{resnet_He_2016_CVPR} to classify them according to their initial domain. If the content is disentangled, the ResNet should not be able to find enough details to classify them and thus perform no better than a random classification. 

\section{Results}
\label{sec:results}
In the following, we present our results and show how our methodological changes in combination with retraining can outperform the evaluated state-of-the-art methods.

\subsection{Quantitative Comparison With State-Of-The-Art Methods} 

We evaluate the outputs $\bar{x}$ of the generator $G$ of each methodology. To do this, we first train our model to disentangle the domain of the evaluated datasets. Then, we transform all three datasets H\textsubscript{O/C/E} into the target domain $O$ by the models generator $G$. In the last step, we evaluate our $UNet_P$ and other quality measures on these transformed datasets. The results are presented in Table \ref{tab:summary_results_base_hippocampus} and Table \ref{tab:summary_results_ablation_hippocampus}.

\begin{table}[htb]
    \centering
    \begin{adjustbox}{max width=8.5cm}
    \begin{tabular}{cccccc}
        \toprule
        \multirow{2}{*}{\textbf{Method}} & \multicolumn{3}{c}{\textbf{Dice} } & \textbf{Average} & \textbf{Classif.} \\ 
         & \textbf{H\textsubscript{O}} $\uparrow$ & \textbf{H\textsubscript{C}} $\uparrow$ & \textbf{H\textsubscript{E}} $\uparrow$ & \textbf{Dice} $\uparrow$ &   \textbf{Score} $\downarrow$ \\ 
        \midrule
        SD-Net \cite{sdnet2019} & \numprint{0.796230501860475}& \numprint{0.676449510039939}& \numprint{0.669189743248554} & \numprint{0.713956585} & \numprint{0.792095289658906} \\ 
        SA-GAN \cite{semantic_aware2018} & - & \numprint{0.739490606636409}& \numprint{0.478596093023746} & \numprint{0.6090433498} & - \\ 
        Dr-GAN \cite{Tran_2017_CVPR} & \numprint{0.666993981087157}& \numprint{0.667175424652306}& \numprint{0.666450220492456} & \numprint{0.6668732087} & \textbf{\numprint{0.328280093845876}} \\ 
        CDD-GAN & \textbf{\numprint{0.807588301051058}} & \textbf{\numprint{0.805663732064964}} & \textbf{\numprint{0.777580434823253}} &  \textbf{\numprint{0.7944511053}} & \numprint{0.504421584551525} \\
        \bottomrule
    \end{tabular}
    \end{adjustbox}
    \caption{Comparison of segmentation and disentanglement performance with the state-of-the-art approaches.}
    \label{tab:summary_results_base_hippocampus}
\end{table}

The $UNet_P$ trained on the H\textsubscript{O} dataset achieves a Dice accuracy of 84.2\% on the test set (see Table \ref{tab:summary_results_ablation_hippocampus}). The domain changes in H\textsubscript{C} decrease the performance by 17.1\% and in H\textsubscript{E} by 11.9\%. By disentangling domain and content, we aim to achieve good segmentation, while keeping the performance very similar for all three datasets.

First, we evaluate the performance of other state-of-the-art methods in Table \ref{tab:summary_results_base_hippocampus}. SD-Net fails to properly transform the images to another domain and produces a very similar output $\bar{x}$ compared to the input $x$. For the dataset H\textsubscript{C} it achieves only a small improvement and therefore does not generalise well. SA-GAN manages a very good domain transfer between the $O$ and $C$ domains and even improves the performance of $UNet_P$ by almost 7\% on H\textsubscript{C}. However, it fails to deal with the deformations and the performance worsens on the H\textsubscript{E} dataset. It only achieves a Dice score of 47.9\%. Dr-GAN manages to separate domain from content almost perfectly. This is reflected in the Dice score, that shows almost identical results of 66.7\% for all three datasets H\textsubscript{O/C/E}. Furthermore, ResNet-18 does not succeed at all classifying the transformed images according to their initial domain. Since there are three datasets, random classification would achieve a score of 33.3\%, which is almost equal to the 32.8\% classification score. Although the segmentation quality of Dr-GAN is worse than without transformation, the disentanglement and domain transfer capability are very good. 

\begin{table}[htb]
    \centering
    \begin{adjustbox}{max width=8.5cm}
    \begin{tabular}{cccccc}
        \toprule
        \multirow{2}{*}{\textbf{Method}} & \multicolumn{3}{c}{\textbf{Dice} } & \textbf{Average} & \textbf{Classif.} \\ 
         & \textbf{H\textsubscript{O}} $\uparrow$ & \textbf{H\textsubscript{C}} $\uparrow$ & \textbf{H\textsubscript{E}} $\uparrow$ & \textbf{Dice} $\uparrow$ &   \textbf{Score} $\downarrow$ \\ 
        \midrule
        UNet\textsubscript{P}  & \textbf{\numprint{0.841510332108924}}& \numprint{0.671151732051255}& \numprint{0.723409944627634} & \numprint{0.7453573363} & \numprint{0.907236960837394} \\ 
        Dr-GAN \cite{Tran_2017_CVPR} & \numprint{0.666993981087157}& \numprint{0.667175424652306}& \numprint{0.666450220492456} & \numprint{0.6668732087} & \textbf{\numprint{0.328280093845876}} \\ 
        (+Ret.) Dr-GAN \cite{Tran_2017_CVPR} & \numprint{0.6968793115047511}& \numprint{0.696706877934904}& \numprint{0.6982611402050327} & \numprint{0.6972824432} & \textbf{\numprint{0.328280093845876}} \\ 
        (¬Ret.) CDD-GAN & \numprint{0.725202318687709} & \numprint{0.713558663354367} & \numprint{0.668235995547367} &  \numprint{0.7023323259} & \numprint{0.504421584551525} \\ 
        CDD-GAN & \numprint{0.807588301051058} & \textbf{\numprint{0.805663732064964}} & \textbf{\numprint{0.777580434823253}} &  \textbf{\numprint{0.7944511053}} & \numprint{0.504421584551525} \\ 
        \bottomrule
    \end{tabular}
    \end{adjustbox}
    \caption{Ablation study of different setups of our proposed methodology.}
    \label{tab:summary_results_ablation_hippocampus}
\end{table}

\subsection{Ablation Studies}

The proposed CDD-GAN (see Table \ref{tab:summary_results_ablation_hippocampus}) improves the Dice score of H\textsubscript{O/C} by 5-6\% and performs similarly well as Dr-GAN for H\textsubscript{E}. The baselines results in Table \ref{tab:summary_results_base_hippocampus} and CDD-GAN show a similar trend in that disentangling the contrast differences is much easier than disentangling elastic deformations.

Finally, we retrain the UNet on the H\textsubscript{O} data transformed with the CDD-GAN generator ${G-CDD}$. The retrained $UNet_R$ achieves a Dice accuracy of 80.8\% on the H\textsubscript{O} test set. The results on the transformed H\textsubscript{C/E} are very similar. $UNet_R$ achieves a Dice score of 80.4\% on H\textsubscript{C} and also manages to reach 77.4\% on H\textsubscript{E}. We also retrain a UNet on Dr-GAN to make sure that this is not the only reason for improvement. Here we can only see an improvement of 3\% for H\textsubscript{O/C/E}. This last step significantly improves the performance without the need for extra labelled data and shows that the transformation by the generator alone is not enough to achieve the best segmentation results. We also need to retrain the UNet to learn to segment on a more generalised representation of the data and therefore achieve more similar results across different data sources. 

\subsection{Quality of the Generated Images}
\begin{figure}[htb]

\begin{minipage}[b]{1.0\linewidth}
  \centering
  \centerline{\includegraphics[width=8.5cm]{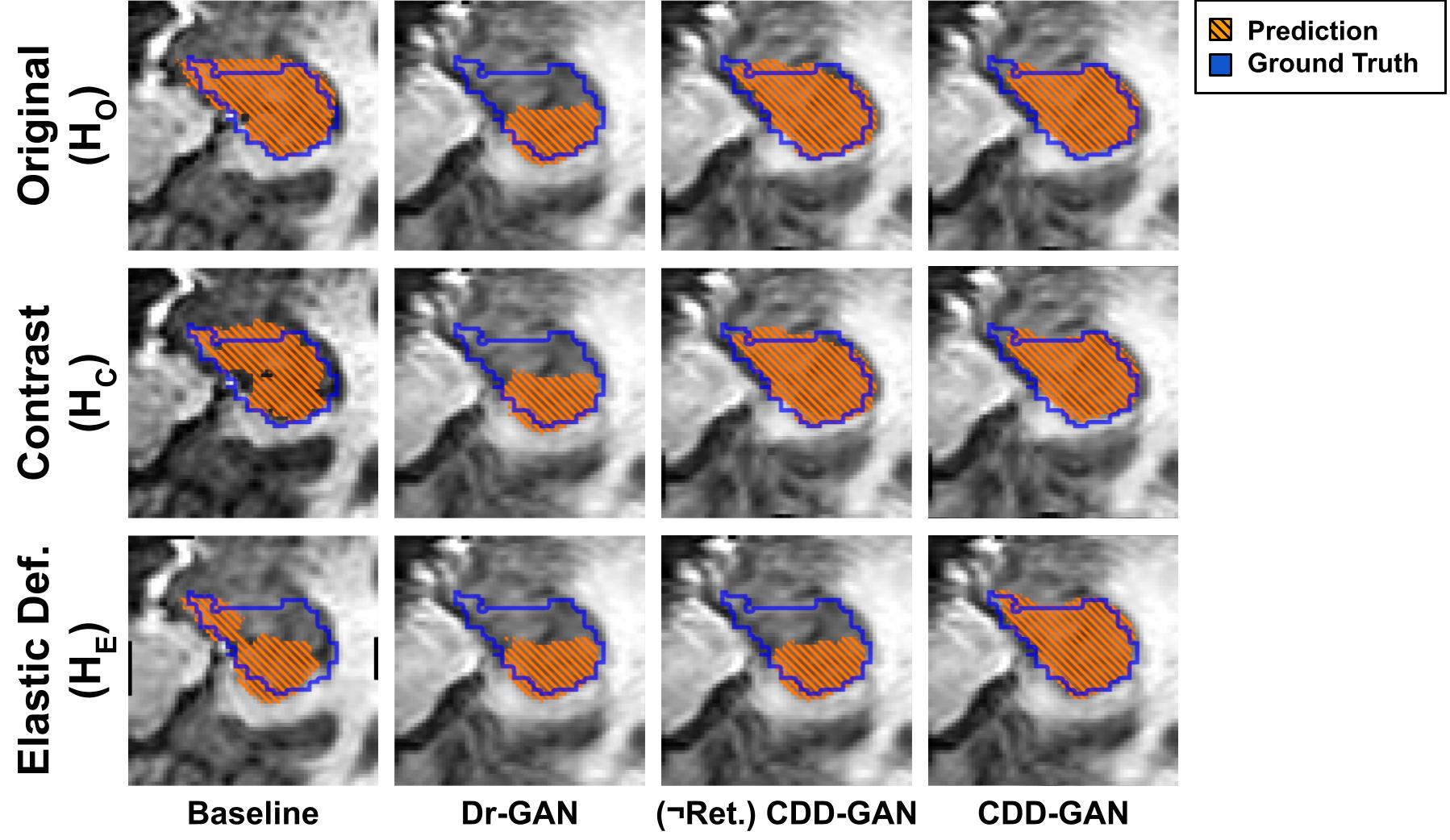}}
\end{minipage}
\caption{Untransformed baseline, transformation results of the generators and the predicted segmentation masks.}
\label{fig:prediction_comparison}
\end{figure}

Figure \ref{fig:prediction_comparison} shows the transformation results of the different generators and the corresponding mask predictions. As can be seen, $UNet_P$ performs very well on the dataset H\textsubscript{O}, since it is the same source as the training data. H\textsubscript{C/E} are much more challenging and show gaps in the mask as well as a complete underestimation of the prediction in H\textsubscript{E}. Dr-GAN achieves almost identical predictions for all three datasets H\textsubscript{O/C/E}, but generally fails to achieve high-quality segmentation results.  

Our proposed CDD-GAN performs well on H\textsubscript{O/C} datasets, with UNet $U_P$ having very similar performance and coming very close to the ground truth in most cases. The biggest problem here is the performance of UNet $U_P$ on H\textsubscript{E}. UNet $U_P$ has difficulty dealing with the differences from H\textsubscript{O/C} and performs very similar to Dr-GAN. 

Using the retrained UNet $U_R$ on the ${G-CDD}$ outputs improves the Dice score even further to 80.8\% and 80.6\% for H\textsubscript{O/C} and 77.8\% for H\textsubscript{E}, which is very close to the segmentation performance achieved by UNet on the original H\textsubscript{O} data. Combining the transformations of CDD-GAN with the more generalised $U_R$ achieves very similar performance for all three datasets, while still predicting high-quality masks and maintaining strong disentanglement. 

\section{Conclusion}
\label{sec:conclusion}
Transferring knowledge between tasks makes it possible to deal with situations in which only limited data is available, as in the segmentation of the hippocampus. We propose the new methodology CDD-GAN to approach this problem. CDD-GAN learns to disentangle the domain differences from the content of the images and generalises the output through its domain-transfer capability. Optimising CDD-GAN on unseen domains in combination with retraining the UNet model with the outputs of the generator network, allows us to improve performance on data from different domains by 6-13\% and outperform the evaluated state-of-the-art methods while maintaining strong disentanglement. All this is achieved without the need for new labelled data, thus opening the way for more robust segmentation models that generalise well to unseen sources even when limited labelled data is available.

\section{Compliance with Ethical Standards}
This research study was conducted retrospectively using human subject data made available in open access by 'Medical Segmentation Decathlon' \cite{medicaldecathlon}. Ethical approval was not required as confirmed by the license attached with the open-access data.

\section{Conflicts of Interest}
No funding was received for conducting this study. The authors have no relevant financial or non-financial interests to disclose.

\bibliographystyle{IEEEbib}
\bibliography{refs}

\end{document}